\documentclass[prl,aps,twocolumn,superscriptaddress,showpacs]{revtex4}
\usepackage{graphicx}
\usepackage{psfrag}
\usepackage{color}
\usepackage{subfigure}
\definecolor{dred}{rgb}{0.7,0.0,0.0}

\begin{document}

\title{      Absence of Hole Confinement in Transition Metal Oxides
             with Orbital Degeneracy
}

\author {     Maria Daghofer }
\affiliation{ Max-Planck-Institut f\"ur Festk\"orperforschung,
              Heisenbergstrasse 1, D-70569 Stuttgart, Germany }

\author {     Krzysztof Wohlfeld }
\affiliation{ Max-Planck-Institut f\"ur Festk\"orperforschung,
              Heisenbergstrasse 1, D-70569 Stuttgart, Germany }
\affiliation{ Marian Smoluchowski Institute of Physics, Jagellonian
              University, Reymonta 4, PL-30059 Krak\'ow, Poland }

\author {     Andrzej M. Ole\'{s} }
\affiliation{ Max-Planck-Institut f\"ur Festk\"orperforschung,
              Heisenbergstrasse 1, D-70569 Stuttgart, Germany }
\affiliation{ Marian Smoluchowski Institute of Physics, Jagellonian
              University, Reymonta 4, PL-30059 Krak\'ow, Poland }

\author {     Enrico Arrigoni }
\affiliation{ \mbox{Institute of
              Theoretical and Computational Physics,
              Graz University of Technology, A-8010 Graz, Austria} }

\author {     Peter Horsch }
\affiliation{ Max-Planck-Institut f\"ur Festk\"orperforschung,
              Heisenbergstrasse 1, D-70569 Stuttgart, Germany }

\date{February 14, 2008}

\begin{abstract}
We investigate the spectral properties of a hole moving in a
two-dimensional Hubbard model for strongly correlated $t_{2g}$
electrons. Although superexchange interactions are Ising-like, a
quasi-one-dimensional coherent hole motion arises due to effective
three-site terms. This mechanism is fundamentally different from
the hole motion via quantum fluctuations in the conventional spin
model with SU(2) symmetry. The orbital model describes also
propagation of a hole in some $e_g$ compounds, and we argue that
orbital degeneracy alone does not lead to hole
self-localization.\\
{[{\it Published in Phys. Rev. Lett.} {\bf 100}, 066403 (2008).]}
\end{abstract}


\maketitle

\begin{figure}[t!]
    \subfigure[]{\includegraphics[width=0.08\textwidth]{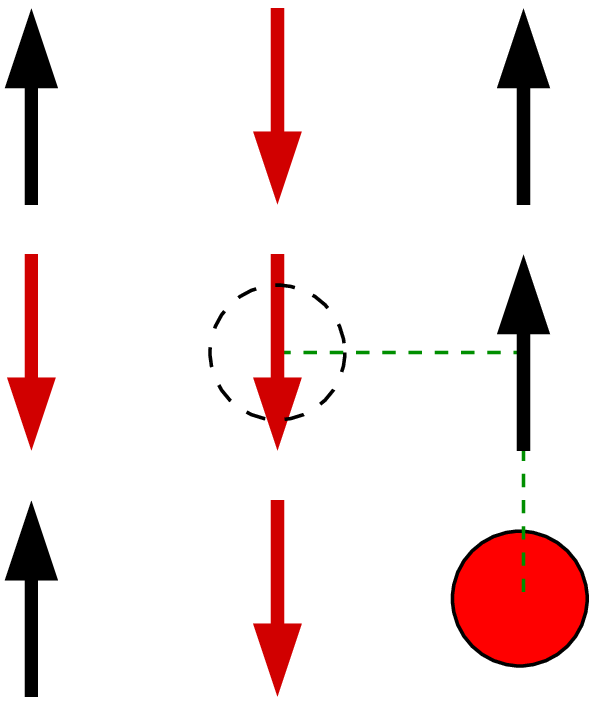}}
    \hspace{0.03\textwidth}
    \subfigure[]{\includegraphics[width=0.08\textwidth]{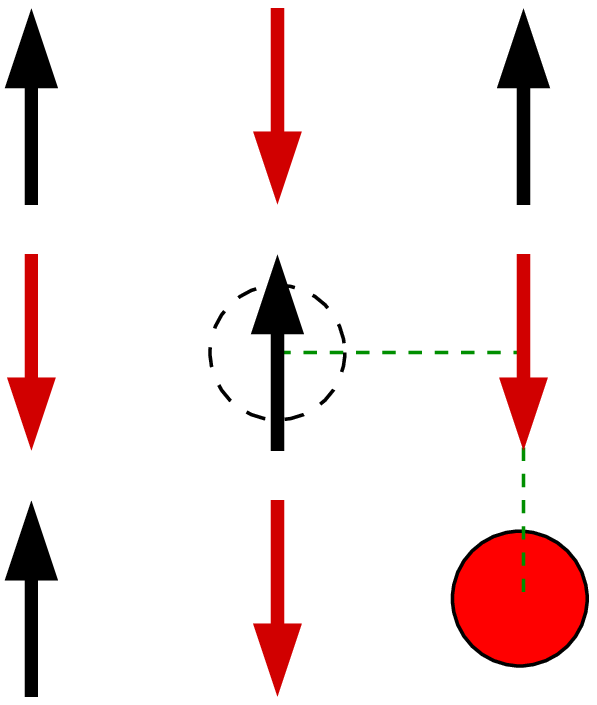}}
    \hspace{0.03\textwidth}
    \subfigure[]{\includegraphics[width=0.1\textwidth]{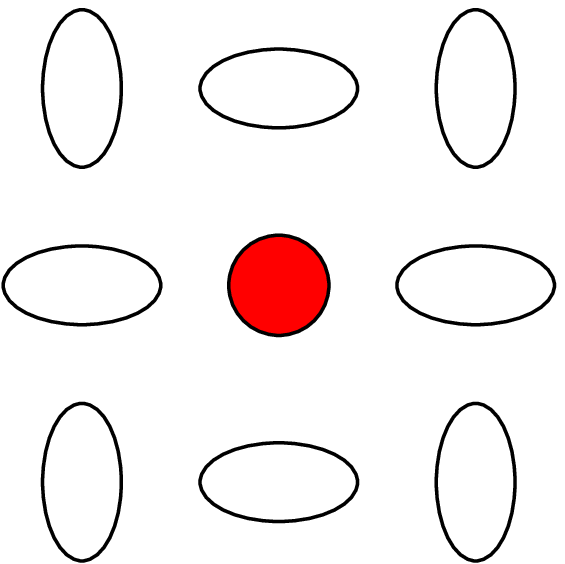}}
    \hspace{0.02\textwidth}
    \subfigure[]{\includegraphics[width=0.1\textwidth]{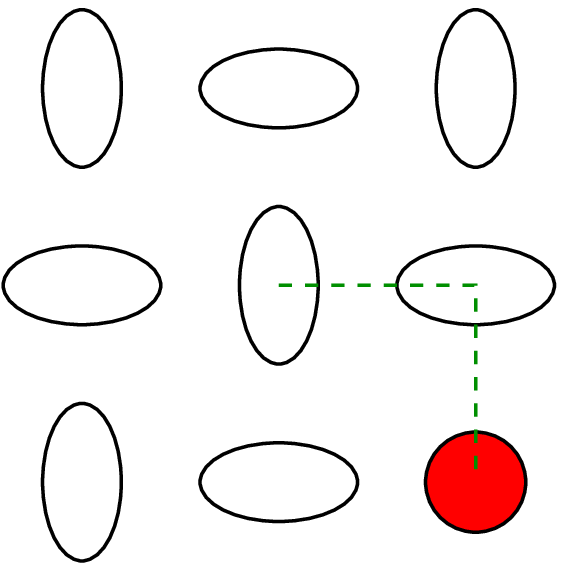}}\\[-1em]
\caption{(color)
Hole motion in:
(a-b) the spin model with SU(2) symmetry, and
(c-d) the $e_g$ orbital model.
Defects in the AF background (a) caused by hole hopping are healed (b)
by spin fluctuations, while for a hole in the AO state (c),
interorbital hopping does not generate defects at all (d).}
\label{fig:hopp_spin_eg}
\end{figure}

One of the fundamental problems in solid state physics consists in
understanding the motion of an electron or hole coupled to the other
degrees of freedom in a material. In many cases, the other degrees
of freedom (spin, orbital or phonon excitations) can increase the mass
of the carrier and possibly localize it.
For example, the undoped parent compound of high-$T_c$ cuprates is an
antiferromagnetic (AF) Mott insulator due to electron-electron
repulsion. A hole doped into it was at first thought to be localized,
because its movement would disturb the AF background and thus cost
energy~\cite{Bri70}. Only two decades later it was found out that
quantum spin fluctuations heal the background and lead to a coherent
hole motion~\cite{Kan89,Mar91}, see Figs.~\ref{fig:hopp_spin_eg}(a-b).
This shows, how important it is to critically assess any approximation
used and to identify possible mechanisms of the coherent hole motion.

Novel aspects of localization occur in \emph{orbital} models.
The superexchange (SE) is then no longer SU(2) symmetric, and the lower
symmetry~\cite{Har03} leads to anisotropy and often to frustrated
interactions~\cite{Tan07}. The resulting variety of possible scenarios
\cite{Zaa93} make compounds with orbital degree of freedom at once very
interesting for material science and challenging to theory. The most
relevant systems of strongly correlated orbitals are (nearly) degenerate
$e_g$ or $t_{2g}$ orbitals. While $e_g$ orbitals are of interest in
colossal magnetoresistance manganites, $t_{2g}$ bands are relevant to,
e.g., cubic titanates~\cite{Kei00} or vanadates~\cite{Ulr03}.

Two $e_g$ orbitals describe the two-dimensional (2D) ferromagnetic
(FM) planes of LaMnO$_3$~\cite{vdB00}, K$_2$CuF$_4$~\cite{Hid83}
or Cs$_2$AgF$_4$~\cite{McL06}. The ground state of the orbital SE
model has Ising-like alternating orbital (AO) order, where quantum
fluctuations are largely suppressed \cite{vdB99}. Nevertheless, a
hole doped into the AO background finds a way to move in {\it
manganites\/}, see Figs. \ref{fig:hopp_spin_eg}(c-d). Here, we
investigate hole motion in a FM plane with two active $t_{2g}$
orbitals (similar to manganites, spin excitations could contribute
for the AF coupling between the FM planes \cite{Bal01}), where AO
order is also Ising-like and interorbital hopping is excluded.
This situation arises, when a crystal field splits $t_{2g}$
orbitals in $d^1$ or $d^2$ systems. If one of the three $t_{2g}$
orbitals is either empty ($d^1$) or fully occupied ($d^2$), the
remaining two can form the AO order, as e.g. in the planes of
Sr$_2$VO$_4$ \cite{Mat05} with possible weak FM order
\cite{Noz91}. In addition to $t_{2g}$ compounds, this "$t_{2g}$
model" also describes $e_g$ orbitals in the above mentioned
fluorides, where a crystal field induces
$d_{z^2-x^2}/d_{z^2-y^2}$-type AO order~\cite{Hid83,McL06}, so
both quantum fluctuations and interorbital hopping are quenched
and cannot generate coherent quasiparticle (QP) propagation
\cite{vdB00}, shown in Fig.~\ref{fig:hopp_spin_eg}.

In this Letter, we show that a hole doped in a state with alternating
$t_{2g}$ orbitals is not confined but finds a way to move
{\it coherently\/} via three-site effective hopping terms arising
from SE, i.e., even in a model with strictly nearest-neighbor hopping.
For the present orbital model, long-range hopping is not expected to
be important, because it is straightforward to verify that:
 (i) the second neighbor hoppings flip the orbital flavor
     \cite{noteph}, so they do not contribute to QP dispersion, while
(ii) the third neighbor hoppings which conserve the orbital flavor are
     considerably smaller than the three-site terms for realistic
     parameters.
These latter SE terms are often neglected \cite{notet1}, but here they
play a central role and determine QP propagation. This finding
contradicts naive expectations of absence of coherent hole motion for
the Ising-like SE in the present $t_{2g}$ orbital model. We investigate
the spectral function by a combination of analytic and numerical
methods to arrive at unbiased conclusions.

\begin{figure}
\begin{minipage}{0.28\textwidth}
  \subfigure[]{
    \includegraphics[width=0.6\textwidth]{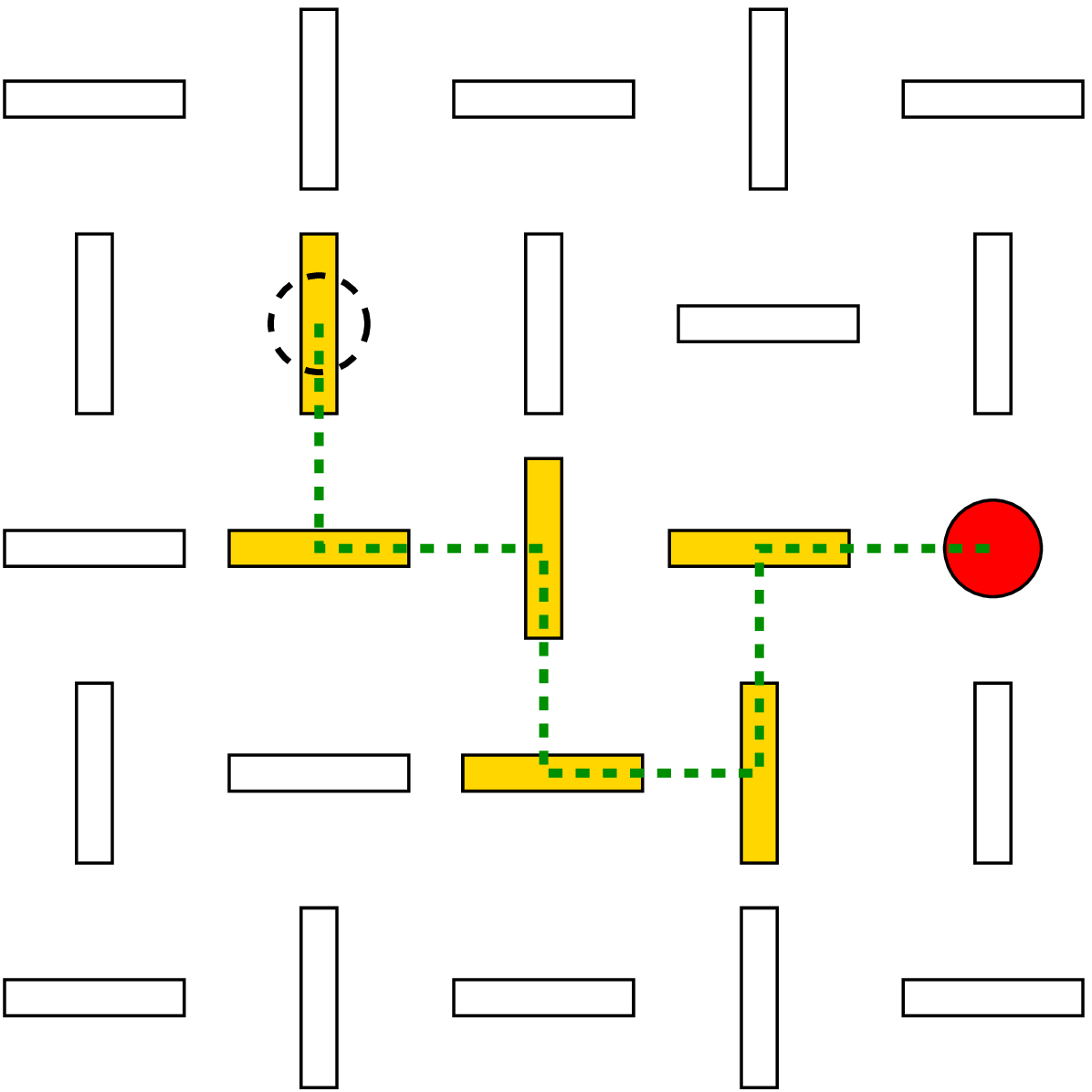}\label{fig:hopp_string}
  }\\[-0.2em]
  \subfigure[]{
    \includegraphics[width=\textwidth]{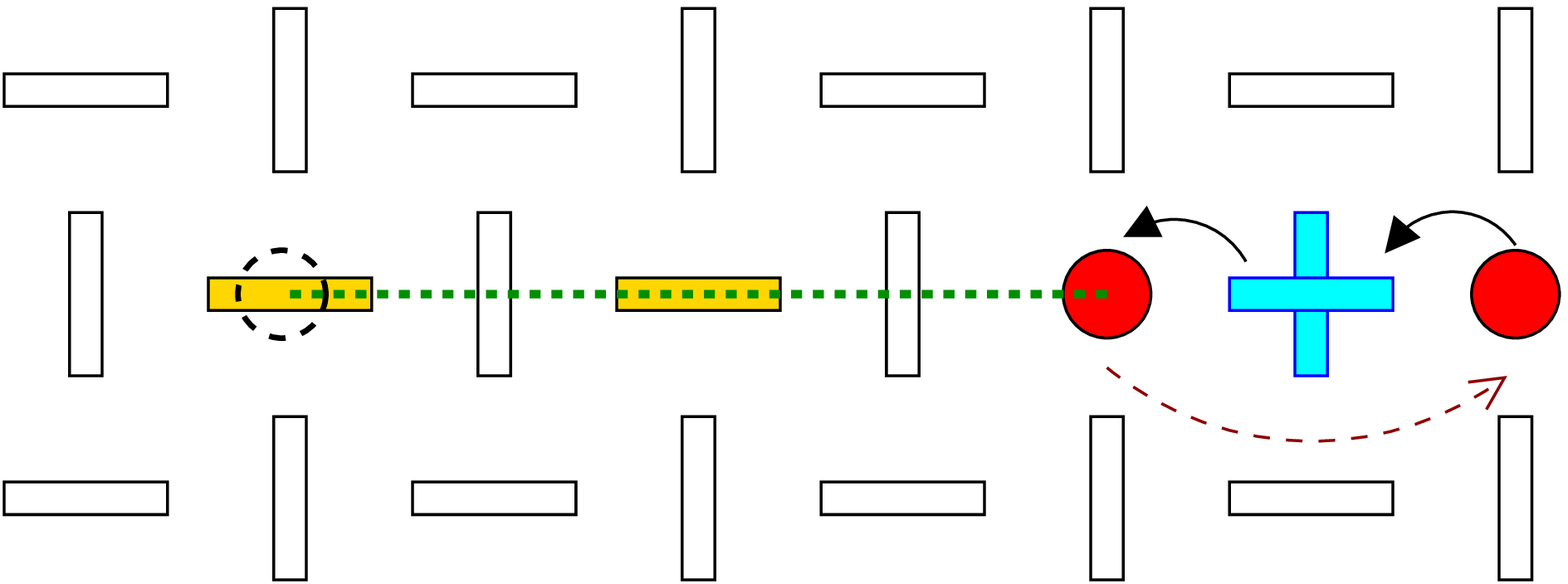}\label{fig:hopp_3site}
  }
\end{minipage}\hskip .7cm
\begin{minipage}{0.11\textwidth}
  \subfigure[]{
    \includegraphics[width=\textwidth]{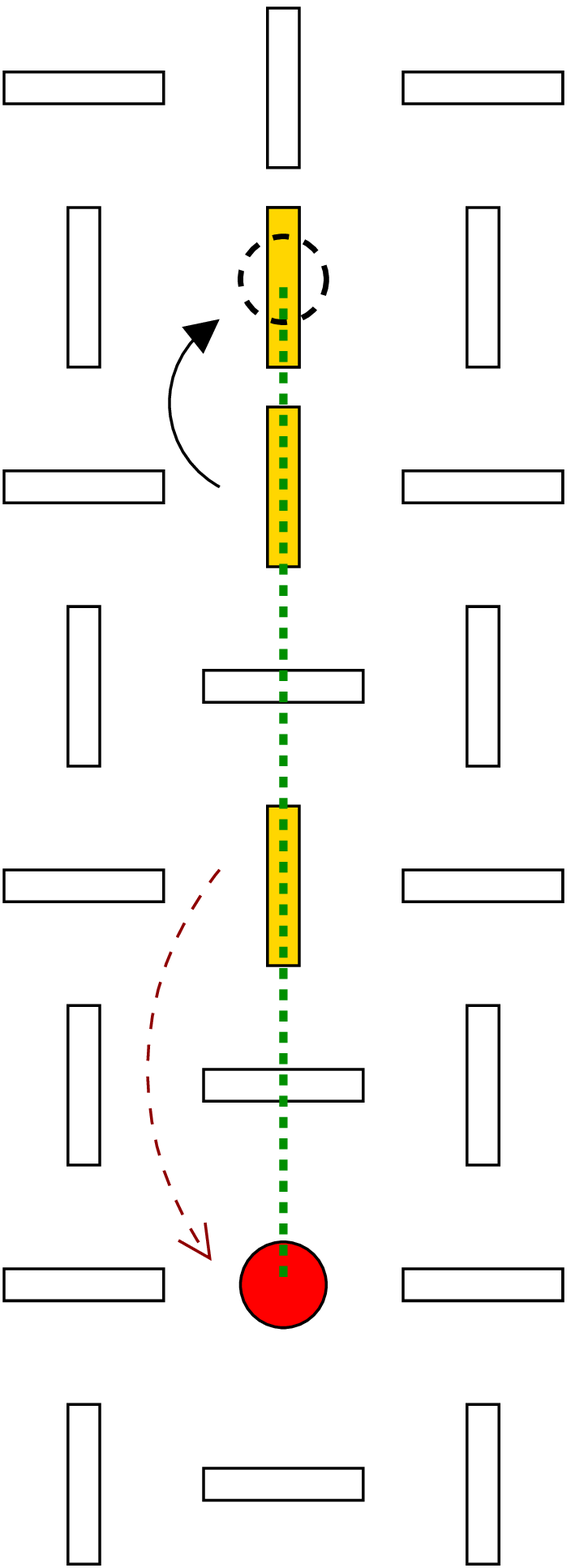}\label{fig:hopp_perp}
  }
\end{minipage}
\caption{(color)
Artist's view of possible hole hopping processes in the orbital model
(\ref{Ht2g}) with AO order ($t\ll U$). Dashed (red) circles indicate
initial (final) positions of the hole. The dotted lines connecting them
give the hole paths, while the shaded (yellow) rectangles mark
electrons that changed orbital due to hole motion. When the hole moves
by hopping $t$ (a), it has to turn by $90^{\circ}$ in each step and
thus creates a zig-zag string of frustrated bonds.
In contrast, in (b) the hole moves along $a$ direction via effective
three-site terms (\ref{H3s0}) generated by intermediate states with two
holes separated by a double occupancy (blue cross). Part (c) shows the
hole moving in the other direction after frustrating three bonds.}
\label{fig:artist}
\end{figure}

\begin{figure}[t!]
\includegraphics[width=6.2cm]{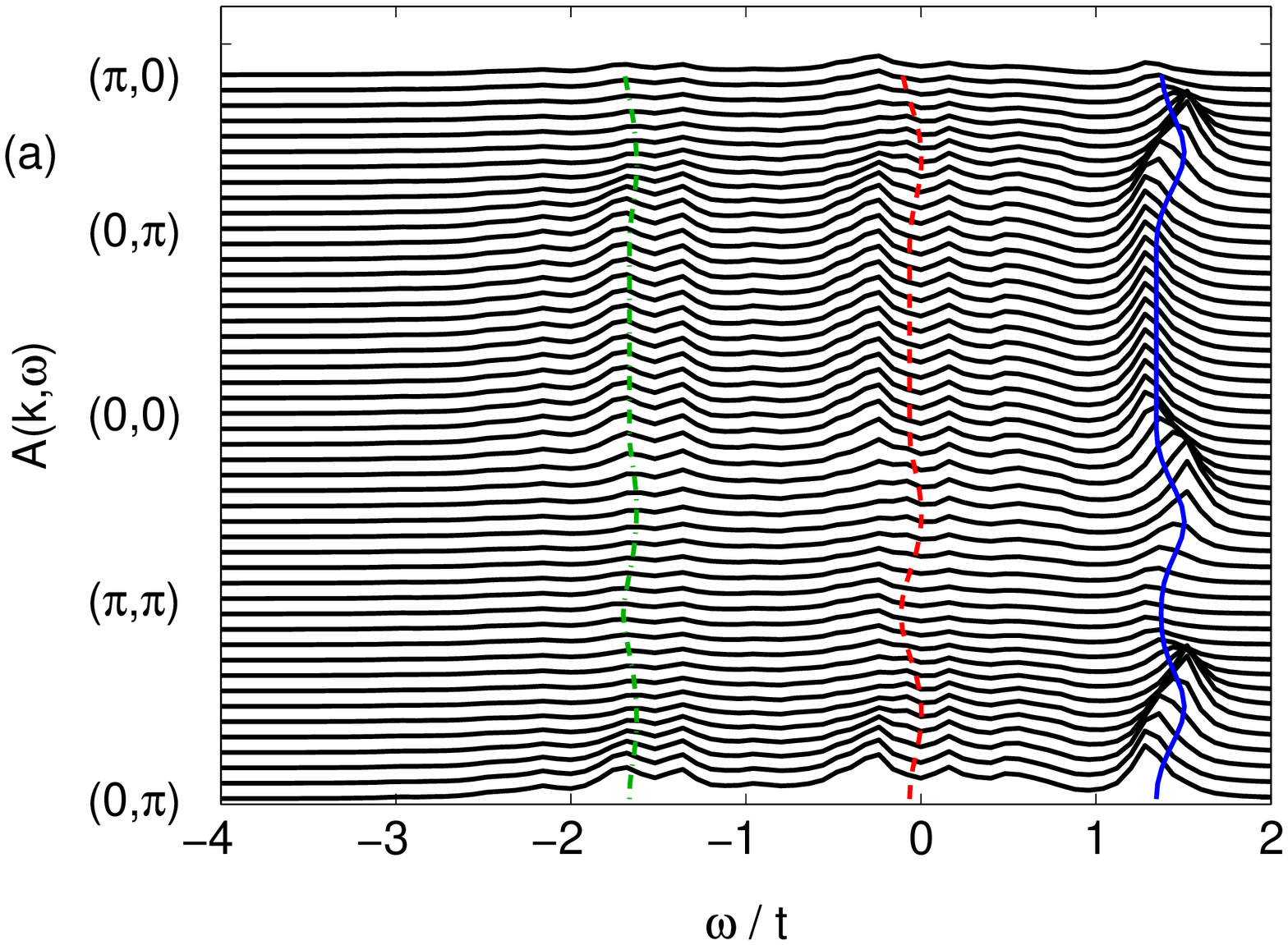}\\[-1.5em]
\includegraphics[width=6.2cm]{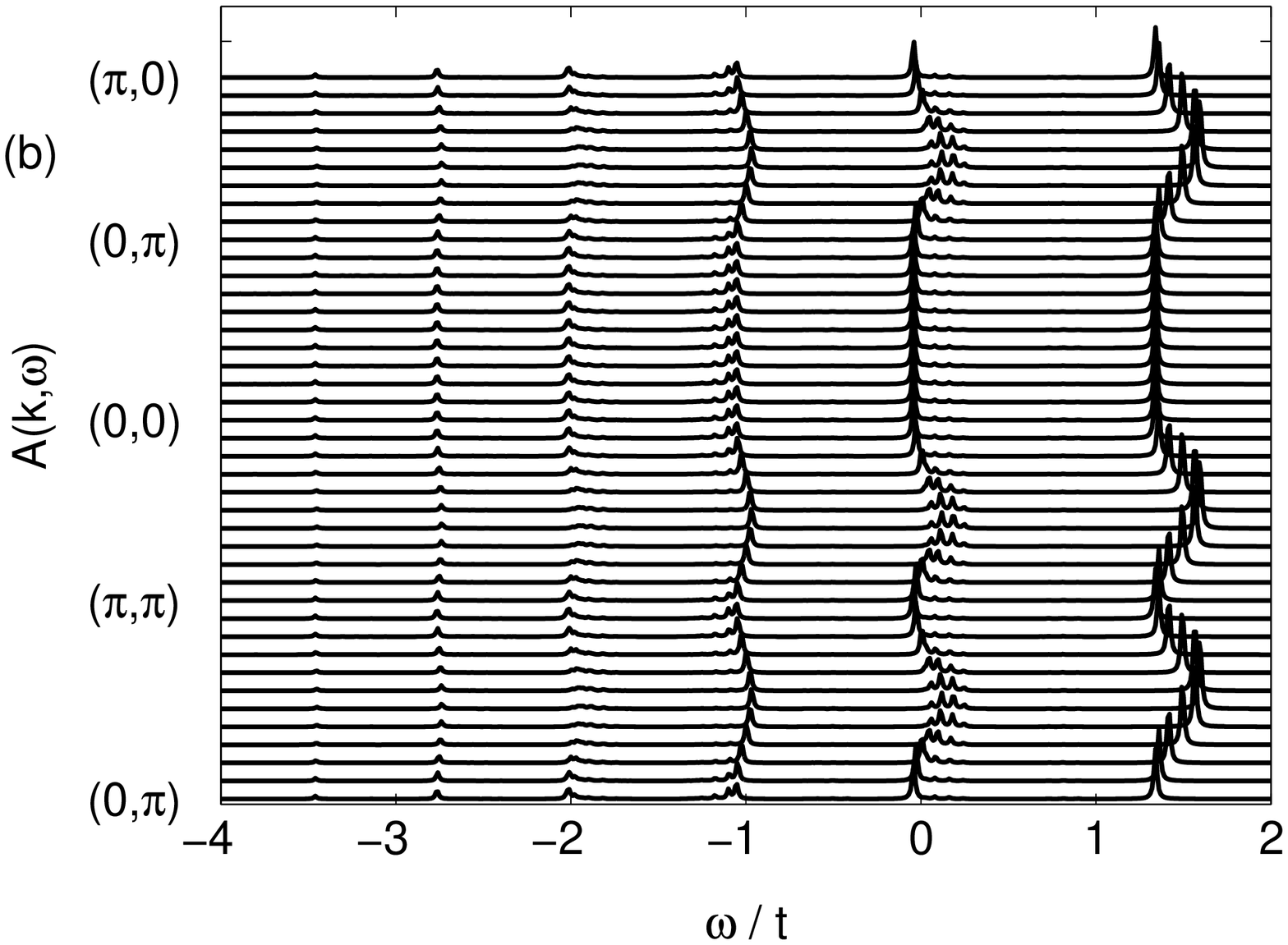}\\[-1em]
\caption{(color online)
Spectral function $A({\bf k},\omega)$ for $b$ orbitals at $U=8t$ along
the main directions of the 2D Brillouin zone obtained for:
(a) the $t_{2g}$ orbital Hubbard model (\ref{Ht2g}) in the VCA with
    a 10-site cluster;
(b) the effective model $\mathcal{H}_{\rm eff}$ (\ref{Heff}) in the SCBA.
First moments of the VCA spectra calculated using the energy intervals
$[0.9t,2t]$, $[-t,0.9t]$, and $[-2.5t,-t]$,
are shown by solid and dashed lines in (a).
}
\label{fig:2d}
\end{figure}

A FM plane with $t_{2g}$ AO order is described by interacting spinless
fermions which undergo one-dimensional (1D) hopping with conserved
orbital flavor:
\begin{equation}
\label{Ht2g} {\cal H}= -t\sum_{
\{ij\}{\parallel}b}a^{\dagger}_{i}a^{}_{j} -t\sum_{
\{ij\}{\parallel}a}b^{\dagger}_{i}b^{}_{j} +U \sum_i n_{ia}
n_{ib}\;.
\end{equation}
Here $a^{\dagger}_{i}$ ($b^{\dagger}_{i}$) creates an electron with
flavor $a$ ($b$) that is allowed to move by hopping $t$ along $b$
($a$) axis and cannot hop along $a$ ($b$) axis in a cubic system
\cite{Har03}; $U$ gives the energy of a doubly occupied site.
Apart from its applicability to $t_{2g}$ and certain $e_g$ orbital
systems, the $t_{2g}$  Hamiltonian (\ref{Ht2g}) is of high theoretical
interest, because it presents --- to our knowledge --- the only
possibility to derive Ising SE from a Hubbard-like model~\cite{ising}.
In the regime of large onsite Coulomb repulsion $U\gg t$, where the
undoped system is a Mott insulator, it reduces to an orbital $t$-$J^z$
model in a similar way as the Hubbard model does reduce to the spin
$t$-$J$ model \cite{Cha77}. The model Hamiltonian,
$\mathcal{H}_{t-J^z}=\mathcal{H}_{t}+\mathcal{H}_{J^z}$, consists of:
\begin{eqnarray}
\label{Ht} \mathcal{H}_{t} &=& -t \sum_{\{ij\}\parallel b}
\tilde{a}^\dag_i \tilde{a}_j -t \sum_{\{ij\}
\parallel a} \tilde{b}^\dag_i \tilde{b}_j\;, \\
\label{HJ}
\mathcal{H}_{J^z} &=& \frac12 J \sum_{\langle ij\rangle }
\left(T^z_i T^z_j - \frac{1}{4}\tilde{n}_i\tilde{n}_j\right)\;.
\end{eqnarray}
Here $\tilde{a}^\dag_i=a_i^\dag(1-n_{ib})$ and
$\tilde{b}^\dag_i=b_i^\dag(1-n_{ia})$ are the creation operators
in the restricted space without double occupancies, as in the spin
$t$-$J$ model \cite{Cha77}, and the sum in Eq. (\ref{HJ}) includes
each bond $\langle ij\rangle$ only once. The corresponding density
operators are $\tilde{n}_{ia}$ and $\tilde{n}_{ib}$, with the
total onsite density $\tilde{n}_i=\tilde{n}_{ia}+\tilde{n}_{ib}$.
$T_i^z=\frac12(\tilde{n}_{ia}-\tilde{n}_{ib})$ stands for the
$z$th component of the pseudospin operator and $J=4t^2/U$ is the
SE energy. A hole inserted into the $b$ orbital of the AO ground
state can move by hopping $t$ only along the bonds $\langle
ij\rangle\parallel a$, i.e., only in one direction, see Fig.
\ref{fig:artist}. This first step costs excitation energy
$E_1=\frac{3}{4}J$, and further steps build a string of orbital
excitations with ever increasing energy [Fig.
\ref{fig:hopp_string}]. In contrast to the spin Hubbard model with
isotropic hopping, the severe restrictions on hole hopping remove
all mechanisms of healing the defects in the AO state --- not only
the quantum fluctuations but even the Trugman loop processes
\cite{Tru88}, which would lead to a coherent propagation, are here
excluded. Consequently, the spectral function $A(\bf{k},\omega)$
for the $t$-$J^z$ model obtained from the self-consistent Born
approximation (SCBA) is independent of momentum ${\bf k}$ and
consists of a ladder spectrum (not shown), with well separated
peaks similar to Refs.~\cite{Kan89,Mar91}. Surprisingly, the
spectral function for the {\it full\/} orbital model (\ref{Ht2g})
obtained within the variational cluster approach (VCA)
\cite{Aic03} exhibits a distinct coherent low-energy
mode~\cite{pes} [shown by a solid line in Fig.~\ref{fig:2d}(a)],
with a 1D dispersion which depends on the orbital.

The VCA is a variational method based on exact diagonalization (ED)
combined with the self-energy functional approach \cite{Pot03},
and has its roots in perturbative cluster approaches for the
Hubbard model~\cite{Mas98}. The method is ideally suited to the present
problem, because it combines unbiased solution of the full Hubbard-like
Hamiltonian (\ref{Ht2g}) on a small cluster (here 10 sites) with access
to the thermodynamic limit. Spontaneous symmetry breaking in ordered
phases is incorporated by optimizing an appropriate `fictitious' field,
in our case a staggered orbital field. Consequently, the resulting
optimal state has almost perfect (Ising-like) AO order. To control our
results, we also performed ED on $4\times4$-site clusters and likewise
found a 1D dispersion with similar bandwidth.

The above results suggest that the $t$-$J^z$ model with its dispersionless
ladder spectrum  cannot reproduce the spectral density of the full
model~(\ref{Ht2g}). The puzzle is resolved by noticing that the complete
low-energy model in second order includes also {\it three-site terms\/}
\begin{equation}
\label{H3s0}
\mathcal{H}_{\rm 3s}^{(0)} \!=
-\frac14 J \sum_{\{imj\}\parallel a} \tilde{b}^\dag_i\tilde{n}_{ma}\tilde{b}_j
-\frac14 J \sum_{\{imj\}\parallel b} \tilde{a}^\dag_i\tilde{n}_{mb}\tilde{a}_j\;,
\end{equation}
where $\{imj\}$ denotes three adjacent sites in a row $\parallel a$
(or a column $\parallel b$) with $m$ in the middle. This effective
hopping term is also obtained perturbatively from Eq. (\ref{Ht2g}) by
allowing one double occupancy next to the inserted hole, and is thus
again of the order $\propto t^2/U$ as the SE term (\ref{HJ}). Figure
\ref{fig:hopp_3site} illustrates how a $b$ electron moves over an
occupied $a$ orbital and interchanges with the hole without affecting
the AO order. This leads to 1D free propagation on two sublattices:
$\varepsilon_{\bf k}^a=\frac12 J\cos(2k_x)$
for the one with occupied $b$ orbitals, and
$\varepsilon_{\bf k}^b=\frac12 J\cos(2k_y)$ for the other.
If we include $\mathcal{H}_{\rm 3s}^{(0)}$
and treat the effective strong-coupling model
\begin{equation}
\label{Heff}
\mathcal{H}_{\rm eff} =
\mathcal{H}_{t}+\mathcal{H}_{J^z}+\mathcal{H}_{\rm 3s}^{(0)}\;,
\label{eq:2}
\end{equation}
the low-energy QP state indeed becomes dispersive similar to the VCA,
both in ED for a $4\times 4$ cluster (not shown) and in the SCBA,
see Fig. \ref{fig:2d}(b). However, the free dispersion $\propto J$
is strongly renormalized, see below. For a single hole, the QP weight
is almost ${\bf k}$-independent in the SCBA. In contrast, in the VCA
for the orbital Hubbard model (\ref{Ht2g}) the spectral weight of the
QP state decreases strongly between ${\bf k}=(0,0)$ and
${\bf k}=(\pi,\pi)$, similar to the decrease of the spectral weight
in the spin Hubbard model \cite{Ste91}. This reflects
the fundamental difference between the Hilbert spaces of the full
(\ref{Ht2g}) and the effective strong-coupling model (\ref{Heff}).

One finds that not only the first, but also all subsequent peaks have
distinct 1D dispersions in the SCBA [Fig.~\ref{fig:2d}(b)].
We attribute this to a 1D propagation analogous to the QP state,
but started after an {\it even\/} number of $t$ hopping steps have
generated string defects in the AO state.
The positions of all peaks evolve from the excitations of the $t$-$J^z$
ladder spectrum where the hole is confined, and the pseudogap between
the QP and the second peak scales as $(J/t)^{2/3}$, similar to the spin
$t$-$J^z$ model \cite{Kan89}. Such processes cannot be properly included
within the VCA using a $10$-site cluster, where too few momenta ${\bf k}$
are available. The dispersive features of Fig.
\ref{fig:2d}(b) are thus replaced by several maxima and the spectral weight
is transferred between them when ${\bf k}$ changes.
Remarkably, the first moments calculated for these structures
(within the relevant energy regimes) show again similar
${\bf k}$-dependence [Fig.~\ref{fig:2d}(a)] as that found in the SCBA.

\begin{figure}[t!]
\includegraphics[width=6.9cm]{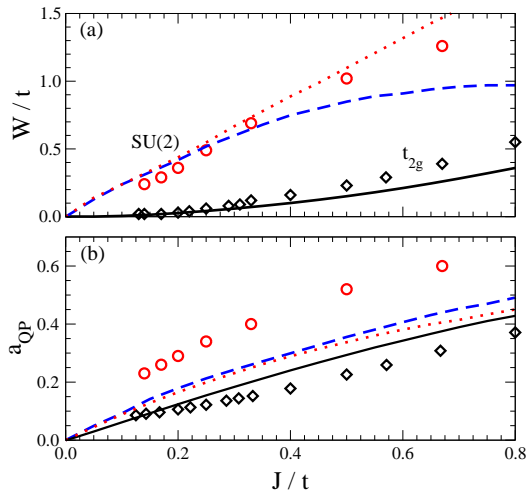}\\[-1em]
\caption{(color online)
QP properties for the $t_{2g}$ orbital ($t_{2g}$) and for the
conventional SU(2)-symmetric spin model [SU(2)] for increasing $J/t$:
(a) bandwidth $W/t$, and
(b) weight $a_{\rm QP}$ averaged over the 2D Brillouin zone.
Diamonds (orbital) and circles (spin) denote the VCA data for the
Hubbard-like models, lines give the SCBA results for the orbital $t$-$J^z$
model with three-site terms (\ref{Heff}) (solid), as well as the usual spin
$t$-$J$ model with (dotted) and without (dashed) three-site terms.
}
\label{fig:qp}
\end{figure}

Figure~\ref{fig:qp} compares QP features of the present $t_{2g}$ model,
obtained within the VCA and the SCBA, with those for the spin
Hubbard/$t$-$J$ model. In the latter case, the bandwidth $W$ at small
$J$ (large $U$) is, as expected, approximately linear in $J$
\cite{Dag94}, both in the VCA (Hubbard model), as well as in the SCBA
($t$-$J$ model), with and without three-site terms. The $t$-$J$ model
with three-site terms gives the complete second-order perturbation
result for the Hubbard model, and consequently its bandwidth agrees
with the VCA data for somewhat larger $J$ (smaller $U$) than the
$t$-$J$ model without three-site terms, see Fig.~\ref{fig:qp}(a).
In the $t_{2g}$ case, where there is no bandwidth $\sim J$ coming from
the quantum fluctuations, the propagation via three-site terms leads
to a bandwidth that is nearly quadratic in $J$. The additional power
arises from the structure of the wavefunction \cite{Hor94}, which
renormalizes the bare three-site bandwidth $\sim J/4$ by an additional
factor $\propto J$ related to the QP weight $a_{\rm QP}$.
It is instructive to take a closer look at the contributions of the
three-site terms to the bandwidth of the usual SU(2) symmetric $t$-$J$
model: The difference between the SCBA results with [dotted line in
Fig.~\ref{fig:qp}(a)] and without (dashed) three-site terms is almost
exactly twice the $t_{2g}$ bandwidth. This factor $2$ is due to the
reduction by $1/2$ of the $t_{2g}$ SE (\ref{HJ}) when compared to the
spin model.

In case of the QP weight $a_{\rm QP}$ [Fig.~\ref{fig:qp}(b)], the
three-site terms have almost no impact on the Hubbard model and
the SCBA results for both models do not differ strongly. While the
VCA gives a similar weight for $t_{2g}$ orbitals as the SCBA, the
values for the spin Hubbard model are considerably larger. Since
the SCBA has been shown to give the same results as quantum Monte
Carlo data for the $t$-$J$ model extrapolated to the thermodynamic
limit~\cite{Bru00}, we believe that the weight given by the VCA is
affected by finite-size effects, which appear to be weaker in the
more classical $t_{2g}$ model.

\begin{figure}[t!]
\includegraphics[width=5.7cm]{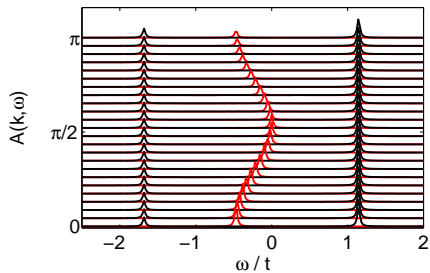}\\[-1em]
\caption{(color online)
Spectral function for the 1D $t_{2g}$ orbital model obtained in the VCA
for $U=8t$. Excitations at $\omega\simeq 1.2t$ and $\omega\simeq -1.7t$
correspond to a hole doped into a localized $a$ orbital,
while a hole doped into a mobile $b$ orbital gives a free propagation
with dispersion $\omega_k=\frac{1}{4}J\cos k+c$.
}
\label{fig:1d}
\end{figure}

Finally, we discuss the 1D chain along $a$ axis [Eq. (\ref{Ht2g})
without the first term], because it gives even clearer insight
into the role played by the three-site terms. In one dimension,
electrons of only one ($b$) flavor can hop, similar to $e_g$
electrons in a 1D FM chain~\cite{Dag04}. Without three-site terms,
one finds just three dispersionless peaks, one arising from a hole
in the mobile orbital and two from the localized orbital
\cite{Dag04}. \emph{With} three-site terms, the hole doped into a
{\it mobile\/} $b$ orbital propagates, see Fig. \ref{fig:1d}. Its
bandwidth is exactly $J$, i.e., we observe the full three-site
dispersion because hole motion is here not renormalized by string
excitations. For a hole doped into a {\it localized\/} $a$
orbital, $t$-$J^z$ and Hubbard-like model give two identical
dispersionless features, which arise from hopping to the sites
next to the initial site of the hole. One might expect the
three-site terms to have some effect, because they allow the hole
to move over the entire chain instead of being confined to just
three sites. However, the energy gain due to this delocalization
is small, and the corresponding spectral feature is invisible in
Fig.~\ref{fig:1d}. For the same reason, processes displayed in
Fig.~\ref{fig:hopp_perp} are not seen in the 2D spectra of
Fig.~\ref{fig:2d} either, where they might be expected to show up
as higher-energy excitations with a dispersion {\it
complementary\/} to that of the first peak.

Summarizing, we have investigated hole motion in a background with
$t_{2g}$ -- and in some cases $e_g$ -- AO order, and have found a
coherent hole motion via three-site terms.
This mechanism is fundamentally different from the ones established
so far in spin systems (quantum fluctuations), or for $e_g$ electrons
in manganites (interorbital hopping). It can also be distinguished from
hole motion via direct longer-range hopping terms because it behaves
differently under particle-hole transformation.
Furthermore, the present model provides a realistic case with purely
classical Ising SE interactions.
As one still finds coherent hole motion, we argue that the hole
confinement and {\it dispersionless \/}ladder excitation spectrum of
the $t$-$J^z$ model --- while being attractive mathematical ideas ---
are {\it never\/} realized in transition metal oxides.

We acknowledge support by the Foundation for Polish Science (FNP),
by the Polish Ministry of Science and Education under Project
No. N202 068 32/1481, and by the Austrian Science Fund
(FWF Project P18505-N16).\\[-2em]


\end{document}